\documentclass[a4paper,12pt]{article}

\usepackage{amssymb,latexsym}

\usepackage{fullpage}

\makeatletter \@addtoreset{equation}{section} 
\makeatother

\begin{document} 
\begin{titlepage}
	\thispagestyle{empty} 
	
	\begin{flushright}
		
		%\hfill{hep-th}\\
		\hfill{DFPD-08TH07}\\
		\hfill{CERN-PH-TH/2008-116} 
	\end{flushright}
	
	\vspace{25pt} 
	\begin{center}
		{ \LARGE{\bf Worldsheet theories for non-geometric string backgrounds }}
		
		\vspace{30pt}
		
		\bf{Gianguido Dall'Agata$^\dagger$ and Nikolaos Prezas$^\flat$}
		
		\vspace{30pt}
		
		{\it $\dagger$ Dipartimento di Fisica ``Galileo Galilei'' $\&$ INFN, Sezione di Padova, \\
		Universit\`a di Padova, Via Marzolo 8, 35131 Padova, Italy}
		
		\vspace{20pt}
		
		{\it $\flat$ Physics Department, \\
		Theory Unit, CERN, \\
		CH-1211, Geneva 23, Switzerland }
		
		\vspace{30pt}
		
		{ABSTRACT} 
	\end{center}
	
	\vspace{10pt} We show that twisted doubled tori can be used to construct a general class of worldsheet models describing non-geometric string backgrounds. 
	By employing a first order formulation of interacting chiral bosons, we first refine the analysis on the general conditions of worldsheet Lorentz invariance and then prove that twisted doubled tori provide good duality symmetric backgrounds. 
	Subsequently we apply our general analysis to several examples which enable us to gain new insight on the difference between geometric, locally geometric and genuine non-geometric backgrounds.

\end{titlepage}

\baselineskip 6 mm

The study of string theory compactifications in the presence of fluxes has dramatically enlarged the number of possible consistent string backgrounds. 
A special role in this context is played by duality symmetries. 
These relations on the one hand  increase further the number of distinct vacua emerging from the effective theories coming from flux compactifications and on the other hand create equivalence classes between different vacua in different models.

One of the most interesting developments in this vein has been the appearance of a new 
class of backgrounds, dubbed ``non-geometric", where non-geometricity implies that at best only a local description in terms of a metric and a rank 2 tensor field is available and that the transition functions between different patches of the compact space contain stringy duality transformations (see \cite{Wecht:2007wu} for a review). 
For this reason, the existence of such backgrounds and their relation to ordinary flux compactifications were first established in the context of their effective 4-dimensional supergravity theories, where the action of duality transformations is best understood. 
However, their 10-dimensional origin or, even worse, their full string theory description
 is problematic, although there is a growing body of evidence that such a description must exist \cite{Hull:2004in,Mathai:2004qc,Bouwknegt:2004ap,Shelton:2005cf,Dabholkar:2005ve,HackettJones:2006bp,Hull:2006va,Shelton:2006fd,Ellwood:2006my,Berman:2007vi,Hull:2007zu,Berman:2007xn,Belov:2007qj,Hull:2007jy,Dall'Agata:2007sr,Berman:2007yf,Pacheco:2008ps,Bianchi:2008cj,Halmagyi:2008dr,Aharony:2008rx}.

An interesting approach to the construction of a consistent worldsheet action for non-geometric backgrounds is that of doubling the number of worldsheet fields corresponding to target space coordinates \cite{Hull:2004in}. 
In this approach the coordinates dual to string winding and momentum modes are treated on equal footing and only a choice of polarization selects the appropriate geometric objects. 
Despite the fact that this ``doubled geometry'' approach has led to some important progress on several aspects of non-geometric backgrounds, like a better understanding of T-fold backgrounds and its quantum equivalence to the standard string formulation \cite{Hull:2006va,Berman:2007vi,Hull:2007zu,Berman:2007xn,Hull:2007jy,Berman:2007yf}, it could be effectively used only for constant background fields or when a  specific dependence on the coordinates is assumed (the so-called T-duality twists).

An alternative action exhibiting an explicit symmetry under T-duality transformations and with doubled target space coordinates had been proposed a while ago in \cite{Tseytlin:1990nb,Tseytlin:1990va}. 
As we will show, this approach allows us to give a generalized geometric description of a quite general class of backgrounds, including a non-trivial dependence on all coordinates (ordinary and dual ones). 
The price one has to pay is the lack of explicit world-volume Lorentz invariance, which, however, can be easily recovered in the class of backgrounds we are going to propose in the ensuing: twisted doubled tori. 

Twisted doubled tori (TDT) as underlying backgrounds of a duality symmetric formulation of string theory have been first proposed in \cite{Dall'Agata:2007sr}, although group manifolds arising from a duality twist with respect to a single coordinate were already introduced in \cite{Hull:2007jy}. 
TDT are local group manifolds with twice as many dimensions as the
usual target space and with a clear action of the duality group O($d,d$), related to the embedding of the adjoint representation of their algebra inside ${\mathfrak o}(d,d)$. 
The special interest in these manifolds was originally motivated by the fact that they provided a unified geometric description of all gauged supergravities, with combinations of ordinary and dual fluxes constrained only by the standard consistency conditions due to the gauging procedure. 
TDT are therefore a clear candidate for providing a stringy origin to arbitrary supergravity models, related to both geometric and non-geometric compactifications.

As already explained in \cite{Dall'Agata:2007sr} and as we will explicitly see in the following, the geometric properties of the target space background are related to the choice of actual spacetime coordinates among those of the TDT. 
In the first order formalism used in this paper, this amounts to deciding which half of the equations of motion for the scalar fields represent constraints for the target space coordinates and which are real equations of motion. 
In the same way one can make contact between this duality symmetric formalism and the ordinary formulation of string theory: one plugs the solutions to the constraint equations in the duality symmetric action, thus recovering an ordinary string $\sigma$-model in terms of a metric and a $B$-field. 
The difference between the geometric and non-geometric case is reflected in the locality of the corresponding $\sigma$-model.

Since in this formalism Lorentz invariance is not granted, it is crucial to prove that backgrounds given by TDT provide good Lorentz invariant $\sigma$-models.
We show that for generic TDT this is indeed the case if one introduces a generalized flux on the group manifold proportional to the group manifold structure constants.
This resembles the Wess--Zumino--Witten construction for strings on group manifolds, although we find that for TDT derived from non-semisimple groups the flux may be trivial.
Specific TDT may also satisfy the Lorentz invariance constraints with several choices
of generalized flux.

An interesting bonus of this formalism is that we can obtain consistent worldsheet theories also in the case of a TDT whose duality matrix $\eta$ which is not constant.
This overcomes some obstacles encountered in \cite{Dall'Agata:2007sr} for the flat group example and further enlarges the possibilities considered in \cite{Hull:2004in}.

We would like to emphasize that the analysis presented here is only a first step towards constructing the conformal invariant worldsheet theories underlying these backgrounds.
Therefore, we will be only discussing the invariance of the action under 
the basic worldsheet symmetries (diffeomorphism invariance, Weyl invariance and local Lorentz rotations) at the classical level.
We furthermore stress that the requirement  of local Lorentz invariance is fundamental in order to have a correspondence with the standard formulation of critical string theory and, as shown in this paper, this already imposes rather non-trivial conditions on the potential doubled geometries.

Good string vacua then correspond to worldsheet actions defining both Weyl and local Lorentz invariant 2-dimensional quantum field theories.
The analysis presented in this paper shows that TDT fulfill the necessary requirement of (on-shell) local Lorentz invariance. 
Furthermore, classical Weyl invariance can be easily proved using the vielbein formalism.
Instead, full quantum consistency, i.e.~quantum conformal invariance, needs to be checked by computing the beta function equations for our TDT backgrounds. This highly non-trivial task is relegated to future work  \cite{inprep} and its importance stems from the fact that it will also yield the spacetime equations of motion and consequently the effective action on the TDT. 

It is natural to expect that this computation will  lead to the equations of motion of gauged supergravity with gauging parameters given by the structure constants of the local group manifold described by the TDT.
Clearly, many of these gauged supergravities will not admit consistent Minkowski or (anti) de Sitter vacua but will only yield runaway potentials.
For this reason one should be aware that the corresponding TDT cannot be in general considered as consistent string theory backgrounds. 
This well-known issue can be addressed by turning on an appropriate dilaton background and several components of the fluxes at the same time, therefore promoting the TDT to a full solution. 
Interestingly enough,  in certain cases these solutions admit an interpretation in terms of a configuration of smeared and intersecting NS5-branes \cite{Ellwood:2006my}. 

In this paper we prefer to focus on simpler TDT which, although they might not all be solutions, provide us with very tractable toy-models for understanding some of the intricacies of non-geometric backgrounds. 
In particular, our ultimate goal would be to demonstrate how a generic
gauged supergravity theory can arise from strings moving on specific background configurations. 
This objective is of the same spirit as compactifications of 10-dimensional supergravity theories on manifolds that are not consistent backgrounds (i.e.~solutions) but which lead to well-defined effective theories with run-away potentials admitting interesting domain-wall or cosmological solutions.  

\medskip

In section \ref{sec:duality_symmetric_formulation_of_string_theory} we revisit the duality symmetric action proposed in \cite{Tseytlin:1990nb,Tseytlin:1990va}, discussing the general equations of motion and Lorentz invariance constraint. 
We also review the action of duality transformations. 
In section \ref{sec:General_solutions} we subsequently discuss various solutions to the Lorentz invariance constraint. 
We first review some known backgrounds, where, for instance, the generalized metric is constant or depends only on half of the doubled coordinates. 
Then we introduce the twisted doubled tori and explicitly show how they realize Lorentz invariance. 
Finally, in section \ref{sec:examples}, we discuss some examples: the flat group, the backgrounds dual to a 3-torus with NS-NS flux and chiral Wess--Zumino--Witten models. 
The analysis of these examples clarifies the role of duality transformations and also their interpretation as geometric, locally geometric or non-geometric.

\medskip

{\bf Note added:} While this paper was in the final stages of preparation we received the preprint \cite{Albertsson:2008gq}, where $D$-branes on doubled tori are studied and a forthcoming paper with a $\sigma$-model description for strings on doubled tori is announced.

\section{Duality symmetric worldsheet theories} 

% (fold)
\label{sec:duality_symmetric_formulation_of_string_theory} 

A natural starting point of a duality symmetric formalism is the doubling of the worldsheet fields corresponding to target space coordinates, including from the very beginning those of the ``ordinary space'' $y^i$ as well as the dual ones $\widetilde y_i$: 
\begin{equation}
	{\mathbb Y}^I = \{y^i , \tilde y_i\}. 
\end{equation}
Once the coordinates have been doubled, it is also quite natural to propose a world-sheet action where the metric and $B$-field are described by a unique generalized metric ${\cal H}$, which is also an element of O($d$,$d$)/O($d$)$\times$ O($d$) 
\begin{equation}
	{\cal H} = \left( 
	\begin{array}{cc}
		g_{ij} - B_{ik}g^{kl}B_{lj} & B_{ik}g^{kj}\\[2mm]
		- g^{ik} B_{kj} & g^{ij} 
	\end{array}
	\right), \label{HgB} 
\end{equation}
so that the string $\sigma$-model is described by an action of the form 
\begin{equation}
	S = \int d {\mathbb Y}^I \wedge \star d {\mathbb Y}^J {\cal H}_{IJ} + \ldots 
\end{equation}
Although this approach is quite natural and both duality symmetry and 2-dimensional Lorentz invariance are manifest, the extra coordinates have to be eliminated through additional constraints to be imposed on the equations of motion and this renders quantization rather complicated. 
Alternatively one could use auxiliary fields, like in \cite{Berman:2007xn}, which however have to be fixed before proceeding further. 
Also, this formalism has been effectively used so far only for constant ${\cal H}$, or for a special dependence on the doubled coordinates, as in \cite{Hull:2007jy}.

The alternative approach we follow in this paper is to temporarily give up 2-dimensional Lorentz invariance and use a manifestly duality symmetric action of interacting chiral bosons \cite{Tseytlin:1990nb,Tseytlin:1990va}. 
This also allows for a clear procedure of getting rid of the dual auxiliary fields through their equations of motion. 
Despite the superficial differences, this approach has been proven to be equivalent to Hull's doubled action for constant background fields \cite{Berman:2007xn}.

The starting action now, not only contains a generalised background metric ${\cal H}_{IJ} = {\cal H}_{JI}$ (although we do not require (\ref{HgB}) yet), but also another metric $\eta_{IJ} = \eta_{JI}$ with $(d,d)$ signature and an antisymmetric 2-tensor $C_{IJ} = - C_{JI}$. 
Moreover, all these background fields in general can depend on all ${\mathbb Y}^I$ coordinates. 
The peculiarity of this action is that it is of first order in worldsheet time derivatives. 
If we generically assume to have worldsheet coordinates $\xi = \{\tau, \sigma\}$, with signature $\{-,+\}$, the $\sigma$-model action reads \cite{Tseytlin:1990va} 
\begin{equation}
	S = \frac12 \int d^2 \xi \left[-\left(C_{IJ}({\mathbb Y})+ \eta_{IJ}({\mathbb Y})\right) \partial_0 {\mathbb Y}^I \partial_1 {\mathbb Y}^J + {\cal H}_{IJ}({\mathbb Y}) \partial_1 {\mathbb Y}^I \partial_1 {\mathbb Y}^J \right]. 
	\label{dualityaction} 
\end{equation}
We would like this action to be Weyl and local Lorentz invariant. 
Invariance under diffeomorphisms can be easily achieved by introducing 2-dimensional vielbeins and worldsheet covariant derivatives everywhere. 
The requirement of on-shell local Lorentz invariance is fundamental in order to have correspondence with the standard formulation of string $\sigma$-models. 
Weyl invariance and local Lorentz invariance are equivalent to the requirement that the trace and the $\epsilon^{ab}$ contraction of the expectation value of the energy-momentum tensor should vanish on-shell. 
The action (\ref{dualityaction}) is not manifestly Lorentz invariant, as time and space worldsheet coordinates are treated on a different footing. 
Only the term depending on $C$ is manifestly invariant, so that demanding local Lorentz invariance yields the following condition 
\begin{equation}
	\eta_{IJ} \left(\partial_0 {\mathbb Y}^I \partial_0 {\mathbb Y}^J + \partial_1 {\mathbb Y}^I \partial_1 {\mathbb Y}^J\right) - 2 {\cal H}_{IJ} \partial_0 {\mathbb Y}^I \partial_1 {\mathbb Y}^J = 0. 
\end{equation}
An extremely useful rewriting of this constraint is the following 
\begin{equation}
	\begin{array}{rcl}
		0 &=& \displaystyle \left(\eta_{IJ} \partial_0 {\mathbb Y}^J - {\cal H}_{IJ}\partial_1 {\mathbb Y}^J\right) \eta^{IK} \left(\eta_{KL} \partial_0 {\mathbb Y}^L - {\cal H}_{KL}\partial_1 {\mathbb Y}^L\right) \\[3mm]
		&& \displaystyle+ \left(\eta- {\cal H}\eta^{-1}{\cal H}\right)_{IJ} \partial_1 {\mathbb Y}^I \partial_1 {\mathbb Y}^J, 
	\end{array}
	\label{Lorentzconstraint} 
\end{equation}
so that Lorentz invariance becomes equivalent to two conditions: ${\cal H}$ has to fulfill
\begin{equation}
	\eta = {\cal H}\eta^{-1}{\cal H}, \label{HOdd} 
\end{equation}
and the $\eta$-norm of 
\begin{equation}
	V_I = \eta_{IJ} \partial_0 {\mathbb Y}^J - {\cal H}_{IJ}\partial_1 {\mathbb Y}^J \label{zeronorm} 
\end{equation}
has to be vanishing, namely 
\begin{equation}
	V_I \eta^{IJ} V_J = 0. 
	\label{nullvector} 
\end{equation}
The condition (\ref{HOdd}) is easy to fulfill by appropriately constructing the ${\cal H}$ matrix, while the norm of $V$ has to vanish on the equations of motion and therefore has to be checked case by case or for classes of backgrounds.

The string equations of motion for a general dependence of the various generalized background fields on the doubled coordinates read 
\begin{equation}
	\begin{array}{l}
		2 \partial_1 \left[\eta_{IJ} \partial_0 {\mathbb Y}^J - {\cal H}_{IJ}\partial_1 {\mathbb Y}^J\right] \\[3mm]
		- 3 \partial_{[I} C_{JK]} \partial_0 {\mathbb Y}^J \partial_1 {\mathbb Y}^K + \partial_I{\cal H}_{JK}\partial_1 {\mathbb Y}^J \partial_1 {\mathbb Y}^K -2 \eta_{JL} \Gamma^L_{IK}(\eta) \partial_0 {\mathbb Y}^J \partial_1 {\mathbb Y}^K=0. 
	\end{array}
	\label{generalequationofmotion} 
\end{equation}
In this equation $\Gamma(\eta)$ are the Christoffel symbols constructed from the  $\eta$ metric and, interestingly, the antisymmetric tensor $C$ appears only through its field strength, so that the equations of motion are invariant under gauge transformations $C \to C + d \Sigma$. 
Besides the various invariances already discussed, the action (\ref{dualityaction}) is also manifestly invariant under the \emph{constant} duality transformations 
\begin{equation}
	{\mathbb Y} \to \Lambda^{-1} {\mathbb Y}, \qquad {\cal H} \to \Lambda^T {\cal H} \Lambda, \qquad C \to \Lambda^T C \Lambda, \qquad \Lambda^T \eta \Lambda = \eta,
	\label{Oddduality} 
\end{equation}
which reduce to constant O$(d,d)$ transformations when $\eta = \Omega$. 
We will see later how to make contact between these transformations, $T$-duality and Buscher's rules.

We stress once more that the action (\ref{dualityaction}) becomes manifestly invariant under diffeomorphisms and classical Weyl transformations if the vielbein formalism is used \cite{Tseytlin:1990va}:
\begin{equation}
	S = \frac12 \int d^2 \xi\,e\, \left[-\left(C_{IJ}({\mathbb Y})+ \eta_{IJ}({\mathbb Y})\right) \nabla_0 {\mathbb Y}^I \nabla_1 {\mathbb Y}^J + {\cal H}_{IJ}({\mathbb Y}) \nabla_1 {\mathbb Y}^I \nabla_1 {\mathbb Y}^J \right]. 
\end{equation}
Full quantum consistency requires the computation of the beta function for generic $C_{IJ}({\mathbb Y})$, $\eta_{IJ}({\mathbb Y}$ and ${\cal H}_{IJ}({\mathbb Y})$ functions.
Only those for which the beta function is vanishing can be taken as consistent string vacua.

% section duality_symmetric_formulation_of_string_theory (end)
\section{General solutions} \label{sec:General_solutions}

Now that we have established the action, its equation of motion and the Lorentz invariance constraint, we present various backgrounds for which the equations of motion (\ref{generalequationofmotion}) imply the Lorentz constraint. 
We start with some simple examples (some already worked out in \cite{Tseytlin:1990va}), which allow us to make contact with the usual $\sigma$-model formulation and Buscher's rules, and then introduce the TDT as new general backgrounds satisfying the Lorentz invariance constraint.

\subsection{Simple vacuum backgrounds} 

% (fold)
\label{sec:simple_vacuum_backgrounds} 

The first easy case we can analyze is that of having \emph{constant background fields}. 
In this case the general equation of motion (\ref{generalequationofmotion}) reduces to 
\begin{equation}
	\partial_1 \left[\eta_{IJ} \partial_0 {\mathbb Y}^J - {\cal H}_{IJ}\partial_1 {\mathbb Y}^J\right] =0, \label{equationofmotionconstants} 
\end{equation}
because $\partial_I \eta_{JK} = \partial_I {\cal H}_{JK} = \partial_I C_{JK} = 0$. 
For closed strings, this condition implies that 
\begin{equation}
	V_I = \eta_{IJ} \partial_0 {\mathbb Y}^J - {\cal H}_{IJ}\partial_1 {\mathbb Y}^J =0 
\end{equation}
and therefore (\ref{zeronorm}) is identically satisfied. 
For constant backgrounds, we can always put the $\eta$ metric in the canonical form by rescaling the ${\mathbb Y}$ fields, namely equal to the constant matrix 
\begin{equation}
	\Omega = \Omega^{-1} = \left( 
	\begin{array}{cc}
		0 & 1_d \\
		1_d & 0 
	\end{array}
	\right). 
	\label{Omega} 
\end{equation}
In this basis it is now obvious that the generalized metric ${\cal H}$ can be written in terms of a metric $g$ and a $B$-field as in (\ref{HgB}) and also that the duality transformations (\ref{Oddduality}) have the usual action in terms of the same fields. 
In particular, for the choice $\Lambda = \Omega$ one exchanges $y^i \leftrightarrow \tilde y_i$ and $g + B \leftrightarrow (g + B)^{-1}$. 
It is also straightforward to show that in this case the equations of motion for the dual coordinates correspond to an explicit form of the Buscher's rules for constant metric and B-field, so that replacing them in the action by their solution to these equations of motion one gets the standard $\sigma$-model 
\begin{equation}
	S = \int d^2\xi \left(\frac12 g_{ij} y^{i\prime} y^{j\prime} - \frac12 g_{ij} \dot{y^i} \dot{y^j} + B_{ij} \dot{y^i} y^{j\prime}\right), 
\end{equation}
where $' \equiv \partial_\sigma$ and $\dot{} \equiv \partial_\tau$. 
When the background fields are constant one can also prove easily \cite{Berman:2007xn} the equivalence of this action to the one proposed in \cite{Hull:2004in}. 
For constant background fields, we are effectively working on a torus and therefore all the ${\mathbb Y}^I$ coordinates parametrize isometries of the background. 
Hence one can safely think of this duality transformations as proper T-dualities.

A simple generalisation of this analysis goes through when ${\cal H}$ depends only on half of the coordinates, for instance the $y^i$ (we still assume $\eta = \Omega$). 
In this case the equations of motion of the dual coordinate fields still read 
\begin{equation}
	\partial_1 \left[\partial_0 y^i - {\cal H}^i{}_{J}\partial_1 {\mathbb Y}^J\right] =0 
\end{equation}
and the Lorentz constraint is still identically satisfied, since it is proportional to the $\tilde y_i$ equations of motion. 
The resulting $\sigma$-model is a standard $\sigma$-model with $y$-dependent couplings $g$ and $B$. 
Due to the special dependence on the various coordinates of the doubled space one can still think of these duality transformations as T-dualities along the coordinates parametrizing isometries of the compact space, these same coordinates appearing always only under differentials.

If one introduces a general dependence on all the ${\mathbb Y}^I$ coordinates, though, Lorentz invariance does not necessarily follows. 
Moreover the duality transformations (\ref{Oddduality}) are not directly related to the standard formulation of T-duality transformations as they may mix coordinates that are not related to isometries of the background space. 
An interesting instance where Lorentz invariance can be successfully and easily implemented is the case of non-trivial ${\cal H}$ and $\eta$, but such that 
\begin{equation}
	{\cal H}(\mathbb Y) = \eta(\mathbb Y). 
\end{equation}
In this case the general equation of motion (\ref{generalequationofmotion}) reduces to 
\begin{equation}
	\partial_1 V_I - \Gamma^L_{IJ}(\eta) V_L \partial_1 {\mathbb Y}^J + \frac32 \partial_{[I}C_{JK]} \eta^{KL} V_L \partial_1 {\mathbb Y}^J= 0, 
\end{equation}
which can also be rewritten using the covariant derivative $\nabla(\eta)$: 
\begin{equation}
	\partial_1 {\mathbb Y}^J \left(\nabla(\eta)_J V_I - \frac32 \partial_{[J}C_{IK]}\eta^{KL} V_L \right) = 0. 
\end{equation}
Contracting this equation with $V^I = V_M \eta^{MI}$ the last term disappears and the equation of motion then implies 
\begin{equation}
	\partial_1 {\mathbb Y}^J V^I \nabla_J V_I = \partial_1 {\mathbb Y}^J \partial_J \left(V^2\right) = \frac12 \partial_1 \left(V^2\right) =0 
\end{equation}
and, for a closed string, this results in a null norm for the vector $V$ with respect to the $\eta$ metric. 
This, together with ${\cal H} = \eta$, which obviously satisfies (\ref{HOdd}), proves the invariance of the action under local Lorentz transformations.

% section simple_vacuum_backgrounds (end)
\subsection{Twisted doubled tori} 

% (fold)
\label{sec:twisted_doubled_tori}

Besides the simple examples shown above, it would be desirable to find a general solution to the Lorentz invariance constraint (\ref{Lorentzconstraint}). 
As already explained, this is very difficult to achieve for generic backgrounds with arbitrary coordinate dependence. 
The strategy attempted in \cite{Tseytlin:1990va} was a perturbative one, around a point where the background fields could be taken constant. 
This, however, did not lead very far.

In this note, instead, we focus on a specific class of backgrounds. 
As mentioned above, Twisted doubled tori constitute an interesting class of candidate string backgrounds for generating arbitrary gauged supergravities as effective theories for the light modes. 
For this reason we give here a constructive proof that the duality symmetric $\sigma$-model for these doubled manifolds is Lorentz invariant on-shell.

TDT can be constructed as (local) group manifolds in the following way. 
One starts by selecting a group representative $g({\mathbb Y}) = {\rm exp}({\mathbb Y}^I {\mathbb X}_I) \in {\cal G}$, where ${\mathbb X}_I$ are the generators of the corresponding gauge algebra ${\mathfrak g} \subset {\mathfrak o}(d,d)$: 
\begin{equation}
	\left[{\mathbb X}_A,{\mathbb X}_B\right] = {\cal T}_{AB}{}^C {\mathbb X}_C. 
\end{equation}
Separating ${\mathbb X}_I = \left\{Z_i,X^i\right\}$, according to their embedding in O($d,d$), we can rewrite ${\mathfrak g}$ also as
\begin{equation}
	\begin{array}{rcl}
	\left[Z_i,Z_j\right] &=& \tau_{ij}^k Z_k + H_{ijk} X^k,\\[2mm]
	\left[Z_i,X^j\right] &=& \tau_{ki}^j X^k + Q^{jk}_i Z_k,\\[2mm]
	\left[X^i,X^j\right] &=& Q^{ij}_k X^k + R^{ijk} Z_k. 
	\end{array}
\end{equation}
Then one extracts the vielbeins ${\mathbb E}^A$ by inspection of the left-invariant Maurer--Cartan form $\Omega = g^{-1}dg = {\mathbb E}^A {\mathbb X}_A$. 
Finally, in order for the theory to be consistently defined on a compact space, one considers only groups such that a left quotient $\Gamma\backslash{\cal G}$ with respect to the compact subgroup $\Gamma = {\cal G}({\mathbb Z})$ is possible. 
When doing so, the doubled vielbeins 
\begin{equation}
	{\mathbb E}^A = U^A{}_I d {\mathbb Y}^I 
\end{equation}
satisfy 
\begin{equation}
	d {\mathbb E}^A = - \frac12 {\cal T}_{BC}{}^A {\mathbb E}^B \wedge {\mathbb E}^C. 
	\label{twisting} 
\end{equation}
Using these vielbeins we can define two different metrics 
\begin{equation}
	{\cal H}_{IJ} = (U^T)_I{}^A \delta_{AB} U^B{}_J, 
\end{equation}
which is O$(d) \times$ O$(d)$ invariant, and 
\begin{equation}
	\eta_{IJ} = (U^T)_I{}^A \Omega_{AB} U^B{}_J, 
\end{equation}
where $\Omega_{AB} = \Omega^{AB} = \left( 
\begin{array}{cc}
	0 & 1 \\
	1 & 0 
\end{array}
\right)$, which is $O(d,d)$ invariant. 
It should be noted that this metric can be put in a constant form only for flat group manifolds \cite{Dall'Agata:2007sr}.
At this point the first constraint necessary to have Lorentz invariance of the $\sigma$-model, namely (\ref{HOdd}), is satisfied by construction. 
To prove that the other is implied by the equations of motion we first rewrite (\ref{generalequationofmotion}) using the compatibility constraint between the Levi--Civita connection constructed from the metric $\eta$ and the spin connection 
\begin{equation}
	\partial_I U^A{}_J - \Gamma_{IJ}^K U^A{}_K + \omega_{IB}{}^A U^B{}_J = 0, 
	\label{compatibility}
\end{equation}
recalling that the spin connection is determined by the group manifold structure (\ref{twisting}). 
It is also useful to define the structure constants with ``curved indices'' 
\begin{equation}
	f_{IJ}{}^K = {\cal T}_{AB}{}^C U^A{}_I U^B{}_J (U^{-1})^K{}_C 
\end{equation}
and recall that 
\begin{equation}
	{\cal T}_{AB}{}^D \Omega_{DC} = - {\cal T}_{AC}{}^D \Omega_{DB}, \label{tauantisym} 
\end{equation}
because the structure constants are in the adjoint of O($d,d$) by construction. 
The latter condition is equivalent to invariance of the metric $\eta$ 
\begin{equation}
	f_{IJ}{}^L \eta_{LK} = - f_{IK}{}^L \eta_{LJ}. 
\end{equation}
At this point we can simplify the equations of motion by rewriting 
\begin{equation}
	\begin{array}{l}
		\partial_I{\cal H}_{JK}\partial_1 {\mathbb Y}^J \partial_1 {\mathbb Y}^K -\partial_I \eta_{JK} \partial_0 {\mathbb Y}^J \partial_1 {\mathbb Y}^K = \\[3mm]
		2 \partial_1 {\mathbb Y}^J \Gamma_{IJ}^L(\eta) {\cal H}_{LK} \partial_1 {\mathbb Y}^K - 2 \partial_1 {\mathbb Y}^J \Gamma_{IJ}^L(\eta) \eta_{LK} \partial_0 {\mathbb Y}^K - \partial_1 {\mathbb Y}^J f_{IJ}{}^L(\eta) {\cal H}_{LK} \partial_1 {\mathbb Y}^K. 
	\end{array}
\end{equation}
This implies that covariant derivatives are reconstructed in the equations of motion: 
\begin{equation}
	2 \partial_1{\mathbb Y}^L \nabla_L(\eta) \left[\eta_{IJ} \partial_0 {\mathbb Y}^J - {\cal H}_{IJ}\partial_1 {\mathbb Y}^J\right] - 3 \partial_{[I} C_{JK]} \partial_0 {\mathbb Y}^J \partial_1 {\mathbb Y}^K - \partial_1{\mathbb Y}^J f_{IJ}{}^K {\cal H}_{KL} \partial_1{\mathbb Y}^L =0. 
\end{equation}
While the terms in brackets reconstruct the same structure we had in previous examples, and this can be easily recast in the form of the null vector condition (\ref{nullvector}), the rest of the equation can be interpreted as a torsion piece only if 
\begin{equation}
	3 \partial_{[I} C_{JK]} = f_{IJ}{}^L \eta_{LK}. 
	\label{dC} 
\end{equation}
In this case the equation of motion reduces to 
\begin{equation}
	2 \partial_1{\mathbb Y}^L \nabla_L(\eta) \left[\eta_{IJ} \partial_0 {\mathbb Y}^J - {\cal H}_{IJ}\partial_1 {\mathbb Y}^J\right] + \partial_1{\mathbb Y}^J f_{IJ}{}^K \left[\eta_{KL} \partial_0 {\mathbb Y}^L -{\cal H}_{KL} \partial_1{\mathbb Y}^L\right] =0, 
\end{equation}
or, in terms of the vector $V_I$ defined in (\ref{zeronorm}) 
\begin{equation}
	2 \partial_1{\mathbb Y}^L \nabla_L(\eta) V_I + \partial_1{\mathbb Y}^J f_{IJ}{}^K V_K =0. 
	\label{almosteq} 
\end{equation}
Although $V_I = 0$ is a solution to the equations of motion (and also of the Lorentz invariance constraint), equation (\ref{almosteq}) may allow for more general solutions.
However, it is now straightforward to show that, after contracting (\ref{almosteq}) with $V_J \eta^{JI}$ and using (\ref{tauantisym}), one obtains 
\begin{equation}
	\partial_1{\mathbb Y}^L \nabla_L(\eta) (V_I \eta^{IJ} V_J) = \partial_1 (V_I \eta^{IJ} V_J) = 0, \label{finaleq} 
\end{equation}
which implies the zero norm condition (\ref{nullvector}) upon using appropriate boundary conditions for a generic TDT background\footnote{These models are also invariant off-shell under modified Lorentz transformations $\delta_L {\mathbb Y}^I  = \delta_{old} {\mathbb Y}^I + \varepsilon \sigma \eta^{IJ} V_J$, which reduce to the standard ones on-shell for the configurations with $V_I = 0$.}.

This derivation requires that (\ref{dC}) admits a well defined solution. 
This can be proved in the case of the TDT by using (\ref{twisting}) 
\begin{equation}
	\begin{array}{rcl}
		dC &=&\displaystyle \frac12 d {\mathbb Y}^I \wedge d {\mathbb Y}^J \wedge d {\mathbb Y}^K \partial_I C_{JK} = \frac16 d {\mathbb Y}^I \wedge d {\mathbb Y}^J \wedge d {\mathbb Y}^K f_{IJ}{}^K \eta_{LK} \\[3mm]
		&=& \displaystyle\frac16 {\mathbb E}^A \wedge {\mathbb E}^B \wedge {\mathbb E}^C {\cal T}_{AB}{}^D \Omega_{DC} = - \frac13 d {\mathbb E}^A \wedge {\mathbb E}^B \Omega_{AB}, 
	\end{array}
\end{equation}
and proving the integrability condition $d^2 C = 0$. 
Although not obvious, this follows by using (\ref{twisting}) and (\ref{tauantisym}): 
\begin{equation}
	d^2 C = \frac{1}{12} {\mathbb E}^A \wedge {\mathbb E}^B \wedge {\mathbb E}^C \wedge {\mathbb E}^D ({\cal T}_{AB}{}^E {\cal T}_{CD}{}^F \Omega_{EF}) = 0 
\end{equation}
where the last equality follows from (\ref{tauantisym}) and the Jacobi identity 
\begin{equation}
	{\cal T}_{[AB}{}^E {\cal T}_{CD]}{}^F \Omega_{EF} = {\cal T}_{F[A}{}^E {\cal T}_{CD}{}^F \Omega_{B]E} = 0. 
\end{equation}

Having a non-trivial $C$, as we will see, implies that the backgrounds obtained by this procedure do not have exactly the same number of units of flux as expected by the TDT geometry alone, but they receive a further contribution from $dC$, which is also proportional to the same structure constants (\ref{dC}).

The general construction discussed above can be modified and simplified if the TDT is a compact group manifold
\begin{equation}
	{\cal G}_1 \times {\cal G}_2 \subset {\rm O}(d) \times {\rm O}(d) \subset {\rm O}(d,d).
\end{equation}
Whenever this is the case the structure constants are not only in the adjoint of O$(d,d)$, but also of O$(d) \times$ O$(d)$.
Hence, not only (\ref{tauantisym}) is true, but also
\begin{equation}
	{\cal T}_{AB}{}^D \delta_{DC} = - {\cal T}_{AC}{}^D \delta_{DB},
\end{equation}
or, in curved indices,
\begin{equation}
	f_{IJ}{}^L {\cal H}_{LK} = - f_{IK}{}^L {\cal H}_{LJ}.
	\label{antisymH}
\end{equation}
Going through the derivation of the equations of motion once more we see that in this case they are equivalent to
\begin{equation}
	\partial_1 V_I - \partial_1 {\mathbb Y}^{L} \Gamma_{IL}^K(\eta)V_K + \frac32 \, \partial_1 {\mathbb Y}^{L} \partial_{[I}C_{LJ]} \partial_0 {\mathbb Y}^{J} = 0,
\end{equation}
because the remaining term with ${\cal H}$ disappears due to (\ref{antisymH}).
We can now explicitly compute $\partial_1 V_I$ as
\begin{equation}
	\partial_1 V_I = (\partial_1 U^A{}_I) V_A + U^A{}_I \partial_1 V_A,
\end{equation}
where $V_A \equiv \Omega_{AB} U^B{}_J (\partial_0{\mathbb Y}^J - \eta^{JK}{\cal H}_{KL} \partial_1 {\mathbb Y}^L)$, and, using (\ref{compatibility}) and (\ref{antisymH}), we obtain
\begin{equation}
	U^A{}_I \partial_1 V_A - \frac12 \partial_1 {\mathbb Y}^J f_{JI}{}^K \eta_{KL} \partial_0{\mathbb Y}^L - \frac32 \partial_1 {\mathbb Y}^J \partial_{[J}C_{IK]} \partial_0 {\mathbb Y}^{K} = 0.
\end{equation}
For this special class of TDT's we can therefore parallelize the connection by choosing 
\begin{equation}
	3 \partial_{[I} C_{JK]} = -f_{IJ}{}^L \eta_{LK},
\end{equation}
which has the opposite sign of (\ref{dC}). By doing so the equations of motion reduce to
\begin{equation}
	\partial_1 V_A = 0,
\end{equation}
which lead, upon using appropriate boundary conditions, to the first order equations
\begin{equation}
	V_I = \eta_{IJ}\partial_0 {\mathbb Y}^J - {\cal H}_{IJ}\partial_1 {\mathbb Y}^J = 0,
\end{equation}
satisfying identically the Lorentz invariance constraint (\ref{nullvector}).

% section twisted_doubled_tori (end)
\section{Examples} 

% (fold)
\label{sec:examples}

We now discuss some classic examples of non-trivial geometric and non-geometric backgrounds: the flat group, the T-duality chain of ${\mathbb T}^3$ with NS-NS flux and the chiral Wess--Zumino--Witten (WZW) models.

The flat group manifold is interesting because, as noticed in \cite{Dall'Agata:2007sr}, its TDT realization gives a non-trivial manifold with a vielbein that cannot be put in the standard triangular form explicitly described by a metric and a $B$-field, unless one introduces a pointwise redefinition of the tangent space basis. 
With the approach proposed in this letter, however, we can show that this TDT correctly reproduces the expected string $\sigma$-model for this compactification. 
We can also see explicitly that the problematic aspects noticed in \cite{Dall'Agata:2007sr} are related to and solved by the non-constant $\eta$ metric.
As we will see in the following, this model is locally equivalent to flat space and therefore it is trivially satisfying the conformal invariance requirement also at the quantum level, therefore providing a good string vacuum.

The second example is the by now classic chain of T-dual backgrounds obtained from a flat 3-torus with 3-form flux on it. 
Several aspects of this chain of dualities have been considered, leading to the interesting remark that the geometry probed on the double dual background is non-commutative and that probed in the fully dual background is non-associative \cite{Mathai:2004qc,Bouwknegt:2004ap,Shelton:2005cf,Shelton:2006fd,Ellwood:2006my,Belov:2007qj}. 
This example allows us to reconstruct in detail the chain of duality transformations as well as understand better the origin of some aspects of the non-geometric fluxes.
It is interesting to point out that in this case the doubled group manifold is flat, but the spacetime backgrounds may have a non-flat metric instead.
Although it is known that such a model cannot provide a consistent string background (at least in a trivial product with Minkowski space), it is an extremely useful toy model from which one can  understand our formalism and methodology in comparison with the previous literature on the subject \cite{Mathai:2004qc,Bouwknegt:2004ap,Shelton:2005cf,Shelton:2006fd,Ellwood:2006my,Belov:2007qj}.

As last example we discuss a theory consisting of two copies of (anti-)chiral WZW models based on the SU(2) group. This model gives an effective theory with a compact gauge group SU(2) $\times$ SU(2) $\subset$ O(3,3) and hence can be described in terms of a TDT with first order equations of motion, following the prescription at the end of section \ref{sec:twisted_doubled_tori}.
From the TDT point of view the gauge algebra involves either $\tau$ and $R$ fluxes or $Q$ and $H$.
This is an instance of a perfectly consistent geometric background, the SU(2) WZW model, that in terms of the TDT description appears to involve the so-called non-geometric fluxes.
In particular, the conformal invariance of the WZW model automatically guarantees
full quantum consistency of the associated worldsheet theory.
Although aspects of this model have already been considered in \cite{Tseytlin:1990va} and \cite{Dabholkar:2005ve}, it is extremely useful to revisit it at the level of the worldsheet theory, also to appreciate the role of the non-constant $\eta$ metric needed to describe general backgrounds.

\subsection{The flat group} 

% (fold)
\label{sub:the_flat_group} 

The flat group is the first example of twisted tori manifolds, namely group manifolds giving rise to non-abelian gauge algebras through a Scherk--Schwarz compactification \cite{Scherk:1979zr}. 
The peculiarity of this group manifold is that it is flat and therefore there is a coordinate system where the metric is the identity and the $B$-field is vanishing, even though global conditions imply a non-trivial gauge structure of the effective theory as well as a truncation of the massless spectrum \cite{Dall'Agata:2005fm}.

Although the flat group is flat as an ordinary group manifold, its TDT realisation is not flat, because it includes the generators dual to the gauge vectors related to the 10-dimensional $B$-field \cite{Dall'Agata:2007sr}. 
The doubled algebra reads 
\begin{equation}
	\left[ {\mathbb X}_I, {\mathbb X}_J \right] = {\cal T}_{IJ}{}^K {\mathbb  X}_K, 
\end{equation}
where ${\mathbb X}_{i} = Z_i$ correspond to the standard Kaluza--Klein generators, ${\mathbb X}^i = X^i$ correspond to the $B$-field gauge transformations and the non-trivial structure constants are 
\begin{eqnarray}
	{\cal T}_{13}{}^2 = -N, \quad {\cal T}_{12}{}^3 = N, \quad {\cal T}_{1}{}^3{}_2 = -N, \\
	{\cal T}_{1}{}^2{}_3 = N, \quad {\cal T}_3{}^2{}_1 = -N, \quad {\cal T}_2{}^3{}_1 = N. 
\end{eqnarray}

Using the procedure outlined in previous sections, we can obtain the doubled vielbeins for this space by constructing the corresponding group manifold using the doubled coordinates 
\begin{equation}
	{\mathbb Y}^I = \left\{y^i,\tilde y_i\right\}=\left\{x,y,z,\tilde x, \tilde y, \tilde z\right\}. 
\end{equation}
The vielbein matrix is 
\begin{equation}
	U^A{}_I = \left( 
	\begin{array}{cccccc}
		1 & & & & &\\
		& \cos (Nx) & \sin (Nx) & && \\
		& -\sin (Nx) & \cos (Nx) &&& \\
		&&& 1 & N z & -N y \\
		&&& & \cos (Nx) & \sin (Nx) \\
		&&& & -\sin (Nx) &\cos (Nx) \\
	\end{array}
	\right) 
\end{equation}
and it is obvious that it is not in a standard triangular form and that it cannot be put in that form by a pure gauge transformation. 
This is also clear from the doubled space metric 
\begin{equation}
	{\cal H} = \left( 
	\begin{array}{cccccc}
		1 & & & & & \\
		& 1 & & && \\
		& & 1 &&& \\
		&&& 1 & N z & -N y \\
		&&& N z& 1 +N^2 z^2 & -N^2 y z \\
		&&& -N y& -N^2 y z & 1 + N^2 y^2 \\
	\end{array}
	\right), 
\end{equation}
which is not in the standard form (\ref{HgB}). 
This model is made compact by the identifications: 
\begin{equation}
	\begin{array}{lclcl}
		\left\{ 
		\begin{array}{l}
			x \sim x+ 1 \\[2mm]
			y \sim y \cos N + z \sin N \\[2mm]
			z \sim -y \sin N +z \cos N \\[2mm]
			\tilde y \sim \tilde y \cos N + \tilde z \sin N \\[2mm]
			\tilde z \sim -\tilde y \sin N +\tilde z \cos N \\[2mm]
		\end{array}
		\right. 
		&& 
		\begin{array}{l}
			\left\{ 
			\begin{array}{l}
				y \sim y+1 \\[2mm]
				\tilde x \sim \tilde x + N \tilde z 
			\end{array}
			\right. 
			\\[10mm]
			\tilde x \sim \tilde x +1\\[2mm]
			\phantom{\tilde z \sim \tilde z +1} 
		\end{array}
		&& 
		\begin{array}{l}
			\left\{ 
			\begin{array}{l}
				z \sim z+1 \\[2mm]
				\tilde x \sim \tilde x - N \tilde y 
			\end{array}
			\right.\\[10mm]
			\tilde y \sim \tilde y +1 \\[2mm]
			\tilde z \sim \tilde z +1 
		\end{array}
	\end{array}
\end{equation}
The crucial difference between this model and those which can be put trivially in the standard form is the non-trivial dependence of the $\eta$ metric on the doubled coordinates ${\mathbb Y}^I$ 
\begin{equation}
	\eta = \left( 
	\begin{array}{cccccc}
		& & & 1 & N z & -N y \\
		& & & &1& \\
		& & &&& 1\\
		1&&& && \\
		N z&1&& && \\
		-N y&&1& && \\
	\end{array}
	\right), 
\end{equation}
which also gives rise to a non-trivial curvature \cite{Dall'Agata:2007sr}. 
As explained in the previous section, the same algebra constrains the form of the antisymmetric 2-form $C$, which in this case we can explicitly solve as 
\begin{equation}
	C = E^1 \wedge \widetilde E_1 = dx \wedge \left(d\tilde x -N y d \tilde z +N z d \tilde y\right). 
\end{equation}
It can be noted that $C$ is globally defined and that therefore the corresponding ``flux'' $dC$ is trivial on the doubled manifold. 
However, after removing the doubled coordinates and obtained the proper string background, the resulting flux is not trivial on the projected space. 
The topological triviality of the generalized flux $C$ is due to the non-semi-simple nature of the TDT Lie algebra.

Plugging all these ingredients in (\ref{dualityaction}), we can explicitly construct the worldsheet $\sigma$-model for this TDT and obtain its equations of motion. 
These read 
\begin{eqnarray}
	- x^{''} + \dot{\widetilde x}^\prime +N\left( \dot{\widetilde y}^\prime z + \widetilde y^\prime \dot z - \dot y \widetilde z^\prime - y \dot{\tilde z}^\prime\right) = 0, \\[2mm]
	- y^{''} + \dot{\widetilde y}^\prime +N {\widetilde z}^\prime \dot x -N \widetilde z^\prime \widetilde x^\prime -N^2 \widetilde z^\prime \widetilde y^\prime z +N^2 y ({\widetilde z}^\prime)^2 = 0, \\[2mm]
	- z^{''} + \dot{\widetilde z}^\prime -N {\widetilde y}^\prime \dot x+N \widetilde y^\prime \widetilde x^\prime -N^2 \widetilde z^\prime \widetilde y^\prime y +N^2 z ({\widetilde y}^\prime)^2= 0, \\[2mm]
	\left(\dot x - \tilde x^\prime -N z \tilde y^\prime +N y \tilde z^\prime\right)^\prime = 0, \\[2mm]
	\left(\dot y - \tilde y^\prime(1+N^2 z^2) +N \dot x z -N \tilde x^\prime z +N^2 y z\tilde z^\prime\right)^\prime = 0, \\[2mm]
	\left(\dot z - \tilde z^\prime(1+N^2 y^2)- N y \dot x +N y \tilde x^\prime +N^2 y z \tilde y^\prime\right)^\prime = 0. 
\end{eqnarray}
We can now make contact with the expected Lagrangian for the $\sigma$-model on an ordinary flat group by interpreting the equations of motion of the dual coordinates $\tilde y_i$ as constraints and plugging their solution back in the original action. 
A crucial point to be noted is that the equations of motion for the dual coordinates result in total space derivatives and therefore, upon choosing appropriate boundary conditions, yield relations between the space derivative of the dual coordinate and the space and time derivatives of the original coordinates. 
This allows us to get rid of the dual coordinates and obtain a geometric description of the resulting $\sigma$-model in terms of an ordinary metric and $B$-field. 
First one can solve the $\tilde x$ equation of motion by 
\begin{equation}
	\tilde x^\prime = \dot x -N z \tilde y^\prime +N y \tilde z^\prime. 
	\label{const1} 
\end{equation}
Using this constraint in the $\tilde y$ and $\tilde z$ equations of motion one finds an easy expression for the derivatives of the dual coordinates: 
\begin{equation}
	\tilde y^\prime = \dot y, \quad \tilde z^\prime = \dot z. 
\end{equation}
This also tells us that the constraint (\ref{const1}) can be solved completely in terms of the ordinary coordinates as 
\begin{equation}
	\tilde x^\prime = \dot x -N z \dot y +N y \dot z. 
\end{equation}
Using altogether these solutions in the original duality symmetric model one gets an effective Lagrangian that reads 
\begin{equation}
	2 \, {\cal L} = - \dot x^2 + y \ddot y + z \ddot z + {x^\prime}^2 + {y^\prime}^2 + {z^\prime}^2, 
\end{equation}
which is equivalent (up to total derivatives) to the Lagrangian for a free string. 
This is indeed the expected local background, with a flat metric and a zero $B$-field, which is further constrained by the global conditions one imposes to get the proper compact space.

Integration of a different set of coordinates leads to T-dual backgrounds, where the role of the geometric fluxes changes. 
Since the geometry probed in this background is locally that of flat space one would expect that these dual backgrounds are also simply a flat ${\mathbb T}^3$. 
As it is clear from the global identifications needed to make the space compact, however, not all coordinates are related to directions on the ${\mathbb T}^3$ that are also good isometries. 
This implies that global obstructions to ordinary T-duality transformations may arise and that non-local aspects may interfere with the simple interpretation of our duality transformations. 
This is especially evident in the case we would keep as coordinates $y,z$ and $\tilde x$, trying to integrate out $x$. 
The $\tilde y$ and $\tilde z$ equations of motion are easily solved leading to a constraint equation for $x$, which reads: 
\begin{equation}
	x'' = \left[\frac{N^2 \dot x(y^2 + z^2) + N z \dot y - y \dot z + \tilde x'}{1+ N^2 y^2 + N^2 z^2}\right]^{{}^{\bullet}}. 
\end{equation}
Clearly this has no direct solution for $x'$ in terms of a single function depending on $y,z$ and $\tilde x$. 
There is, however, a field redefinition mixing $x$ and $\tilde x$ which solves it identically 
\begin{equation}
	x' = \dot w, \qquad \tilde x' - \dot x N^2 \left(y^2 + z^2\right) = y \dot z - z \dot y +\left(1+N^2 y^2 + N^2 z^2\right)w'. 
	\label{fieldred} 
\end{equation}
Once these substitutions are used throughout, the equations of motion reduce to 
\begin{eqnarray}
	&& -\ddot w + w'' = 0, \\[2mm]
	&& -\ddot y + y'' + N^2 y I^2 + N z \dot I + 2 N \dot z I =0,\\[2mm]
	&& -\ddot z + z'' + N^2 z I^2 - N y \dot I - 2 N \dot y I =0, 
\end{eqnarray}
for 
\begin{equation}
	I = \int d\sigma (w'' - \ddot w). 
\end{equation}
Although we explicitly get flat space once more, it is clear that the field redefinitions (\ref{fieldred}) imply a non-local dependence of the new background coordinate on the dual one.

% subsection the_flat_group (end)
\subsection{The $H$, ${\tau}$, $Q$, $R$ flux chain} 

% (fold)
\label{sub:the_h_tau_q_r_flux_chain} 

The algebra dual to a compactification on a flat 3-torus with non-trivial $H$-flux is summarized by the structure constants 
\begin{equation}
	{\cal T}_{12}{}_3 = -N, \quad {\cal T}_{23}{}_1 = -N, \quad {\cal T}_{31}{}_2 = -N. 
\end{equation}
Using the procedure outlined above we can construct the TDT with vielbein 
\begin{equation}
	U^A{}_I = \left( 
	\begin{array}{cccccc}
		1 & & & & &\\
		& 1 & & && \\
		& & 1 &&& \\
		&-N z&N y& 1 & & \\
		N z&&-N x& & 1 & \\
		-N y&N x&& & &1 \\
	\end{array}
	\right) 
\end{equation}
and generalized metric 
\begin{equation}
	{\cal H} = \left( 
	\begin{array}{cccccc}
		1 +N^2 y^2 +N^2 z^2&-N^2 xy &-N^2 xz & &N z &-N y \\
		-N^2 xy & 1 +N^2 \left(y^2+z^2\right)& -N^2 yz&-N z &&N x\\
		-N^2 xz&-N^2 yz & 1 +N^2 \left(x^2+y^2\right)&Ny&-Nx& \\
		&-Nz&Ny& 1 & & \\
		Nz&&-Nx& & 1 & \\
		-Ny&Nx&& & & 1 \\
	\end{array}
	\right). 
\end{equation}
For this class of examples the $\eta$ metric is completely trivial \cite{Dall'Agata:2007sr} 
\begin{equation}
	\eta = \left( 
	\begin{array}{cccccc}
		& & & 1 & & \\
		& & & &1& \\
		& & &&& 1\\
		1&&& && \\
		&1&& && \\
		&&1& && \\
	\end{array}
	\right), 
\end{equation}
which allows the rewriting of ${\cal H}$ in terms of a $g$ and a $B$ field as in (\ref{HgB}). 
The standard construction of a TDT duality symmetric worldsheet model foreseen in the previous sections however includes a non-trivial 2-form $C$, which, for the algebra at hand reads 
\begin{equation}
	C = \frac13 \left(E^1 \wedge \widetilde E_1+E^2 \wedge \widetilde E_2+E^3 \wedge \widetilde E_3\right). 
\end{equation}
Again, this non-trivial $C$ implies that the actual $B$-field appearing in the string $\sigma$-model after integrating out the dual coordinates is shifted with respect to the one that could be read directly from ${\cal H}$ using (\ref{HgB}).

Once we have constructed the world-sheet Lagrangian (\ref{dualityaction}) from the ingredients described above, the equations of motion read 
\begin{eqnarray}
	- x^{''}(1+N^2 y^2+N^2 z^2) + \dot{\widetilde x}^\prime -2N^2 x' y' y + 2N^2 x (y')^2 +N^2 x y y'' -N \widetilde y'' z -N y' \dot z \nonumber \\
	+N \dot y z^\prime- 2N \widetilde y' z' - 2N^2 z x' z' + 2 N^2 x (z')^2 +N^2 x z z'' +N 2 y' \widetilde z' + N y \tilde z'' = 0, \label{xeom}\\[2mm]
	- y^{''}(1+N^2 x^2+N^2 z^2) + \dot{\widetilde y}^\prime -2N^2 y' z' z + 2N^2 y (z')^2 +N^2 y z z'' -N \widetilde z'' x -N z' \dot x \nonumber \\
	+N \dot z x^\prime - 2N \widetilde z' x' - 2N^2 x y' x' + 2N^2 y (x')^2 +N^2 y x x'' + 2N z' \widetilde x' +N z \tilde x'' = 0, \label{yeom}\\[2mm]
	- z^{''}(1+N^2 y^2+N^2 x^2) + \dot{\widetilde z}^\prime -2N^2 z' x' x + 2N^2 z (x')^2 +N^2 z x x'' -N \widetilde x'' y -N x' \dot y \nonumber \\
	+ N\dot z y^\prime - 2N \widetilde x' y' - 2N^2 y z' x' + 2N^2 z (x')^2 +N^2 z y y'' + 2N x' \widetilde y' + N x \tilde y'' = 0, \label{zeom}\\[2mm]
	\left(\dot x - \tilde x^\prime +N z y^\prime -N y z^\prime\right)^\prime = 0, \label{xteom}\\[2mm]
	\left(\dot y - \tilde y^\prime +N x z^\prime -N z x^\prime\right)^\prime = 0, \label{yteom}\\[2mm]
	\left(\dot z - \tilde z^\prime +N y x^\prime -N x y^\prime\right)^\prime = 0. 
	\label{zteom} 
\end{eqnarray}
Just like in the previous example we have three equations that are total space derivatives and three equations that cannot be put in this form.

\subsubsection{$H$-flux} % (fold)
\label{sub:h_flux}

The first thing is to make contact with the original geometric model, with $H$-flux and flat metric. 
This can be achieved by integrating out the $\tilde y_i$ coordinates and it is fairly easy to do because their equations of motion are total space derivatives. 
The solutions express the dual coordinates completely in terms of the geometric ones: 
\begin{eqnarray}
	\tilde x^\prime = \dot x +N z y^\prime -N y z^\prime, \label{solxt}\\[2mm]
	\tilde y^\prime = \dot y +N x z^\prime -N z x^\prime, \label{solyt}\\[2mm]
	\tilde z^\prime = \dot z +N y x^\prime -N x y^\prime. 
	\label{solzt} 
\end{eqnarray}
Plugging this solution into the original Lagrangian gives the expected effective $\sigma$-model 
\begin{equation}
	\begin{array}{rcl}
		2 {\cal L}_{eff} &=& (x')^2+(y')^2+(z')^2-(\dot x)^2-(\dot y)^2-(\dot z)^2 \\[2mm]
		&& \displaystyle + \frac43 N \,y \left(\dot x z' - \dot z x'\right) + \frac43 N\, x \left(\dot z y' - \dot y z'\right) + \frac43 N \,z \left(\dot y x' - \dot x y'\right), 
	\end{array}
	\label{Hflux} 
\end{equation}
up to boundary terms 
\begin{equation}
	-\frac23 \left(x \dot x + y \dot y + z \dot z\right)\dot{}+ \frac23 \left(x \dot{\widetilde x}+ y \dot{\widetilde y}+z \dot{\widetilde z}\right)' . 
\end{equation}
The $\sigma$-model described by (\ref{Hflux}) can be interpreted as that of a string moving in an ordinary background given by a flat metric and a non-trivial $B$-field 
\begin{equation}
	ds^2 = dx^2 + dy^2 + dz^2, \quad B = \frac23 N\,\left(x dy \wedge dz + y dz \wedge dx + z dx \wedge dy\right) \label{Hbackground} 
\end{equation}
leading to 
\begin{equation}
	H = dB = 2 N \, dx \wedge dy \wedge dz. 
\end{equation}

% subsection h_flux (end)

\subsubsection{Geometric flux} % (fold)
\label{sub:geometric_flux}

A different choice of constraint equations, equivalent to a different choice of coordinates to be considered the dual ones, gives rise to dual backgrounds. 
For instance, one could consider one $T$-duality exchanging the role of the $z$ and $\tilde z$ coordinates, so that the final geometry should be described in terms of $x, y$ and $\tilde z$. 
This means that now we can still replace $\tilde y'$ and $\tilde x'$ using (\ref{solxt}) and (\ref{solyt}), but now (\ref{zteom}) should be interpreted as a real equation of motion, while the constraint equation to solve is the equation of motion for $z$ (\ref{zeom}). 
The latter, however, also becomes a total space derivative, once the constraint equations for $\tilde y$ and $\tilde z$ have been used: 
\begin{equation}
	\left(z' + N \dot x y -N x \dot y - \dot{\tilde z}\right)' = 0. 
\end{equation}
We can therefore proceed again to solve it by appropriately choosing the boundary conditions as 
\begin{equation}
	z' = \dot{\tilde z} -N \dot x y +N x \dot y. 
\end{equation}
The effective Lagrangian finally reads (up to boundary terms) 
\begin{equation}
	{\cal L}_{eff} = (x')^2+(y')^2+(\tilde z'+N x y' -N y x')^2-(\dot x)^2-(\dot y)^2-(\dot {\tilde z} +N x \dot y -N y \dot x)^2 
\end{equation}
and again it is the appropriate Lagrangian for a string moving in a background with zero $B$-field and a non-trivial metric 
\begin{equation}
	ds^2 = dx^2 + dy^2 + \left(d\tilde z +N x dy -N y dx\right)^2, \quad B = 0. 
	\label{taubackground} 
\end{equation}
This corresponds to a background with a purely geometric flux $\tau$ which is the appropriate T-dual of (\ref{Hbackground}): 
\begin{equation}
	d e^3 = d \left(d\tilde z +N x dy -N y dx\right) = 2 N dx \wedge dy. 
\end{equation}

% subsection geometric_flux (end)

\subsubsection{$Q$-flux} % (fold)
\label{sub:_q_flux}

A series of two T-dualities is equivalent to integrating out two of the three original coordinates. 
In the following we use as coordinates $x$ and $\tilde y, \tilde z$ (or their combinations) and use as constraints the equations of motion for $\tilde x$, $y$ and $z$: (\ref{xteom}), (\ref{yeom}) and (\ref{zeom}). 
The constraint equation for $\tilde x$ is solved exactly like in the previous cases by (\ref{solxt}). 
This time, however, after plugging this solution in the equations of motion for $y$ and $z$ we don't get simple total space derivatives equations. 
The constraint equations now read 
\begin{eqnarray}
	\left(\tilde y' +N z x' -N x z'\right)\dot{}-2N x' \left(\tilde z' +N x y' -N y x'\right) -N x \left(\tilde z' +N x y' -N y x' - \dot z\right)' \nonumber\\[2mm]
	- y'' + 2N \dot x z' =0, &&\label{consty} \\[2mm]
	\left(\tilde z' +N x y' -N y x'\right)\dot{}+2N x' \left(\tilde y' +N z x' -N x z'\right) +N x \left(\tilde y' +N z x' -N x z' - \dot y\right)' \nonumber\\[2mm]
	- z'' - 2N \dot x y' =0. 
	&&\label{constz} 
\end{eqnarray}
Because of this structure it is clear that it is not possible to give a simple local expression of the $y$ and $z$ fields in terms of the dual ones. 
We can also notice that part of the constraint equations (\ref{consty}) and (\ref{constz}) are proportional to the $\tilde y$ and $\tilde z$ equations of motion 
\begin{equation}
	\begin{array}{l}
		\left(\tilde y' +N z x' -N x z'\right)' - \dot{y'} = 0, \\[2mm]
		\left(\tilde z' +N x y' -N y x'\right)' - \dot{z'} = 0. \\[2mm]
	\end{array}
\end{equation}
We can therefore try to solve these constraint equations modulo the resulting equations of motion, so that we do not affect the final result. 
If we do so, the constraint equations become 
\begin{eqnarray}
	\left(\tilde y' +N z x' -N x z'\right)\dot{}-2N x' \left(\tilde z' +N x y' -N y x'\right) - y'' + 2N \dot x z' = \nonumber \\[2mm]
	=N \alpha x \left(\tilde z' +N x y' -N y x' - \dot z\right)',&&\label{consty1} \\[2mm]
	\left(\tilde z' +N x y' -N y x'\right)\dot{}+2N x' \left(\tilde y' +N z x' -N x z'\right) - z'' - 2N \dot x y' =\nonumber\\[2mm]
	=-N \alpha x \left(\tilde y' +N z x' -N x z' - \dot y\right)'.&&\label{constz1} 
\end{eqnarray}
We now try to satisfy these constraints, by starting from (\ref{consty1}). 
We can collect three types of terms, proportional to $x$, $\dot x$ and $x'$, if we perform the following redefinitions 
\begin{equation}
	\tilde y' +N z x' -N x z' = w' +NxA, \quad y' = \dot w +N x B, \quad \tilde z' +N x y' -N y x' = C, \quad z' = D. 
\end{equation}
These positions are also justified by the fact that $\tilde y$ and $\tilde z$ have to be proportional to the ``geometric coordinate'' and that $y'$ and $z'$, being the dual ones, should be proportional to the time derivative of the gometric ones. 
After these replacements (\ref{consty1}) becomes 
\begin{equation}
	\dot x A + x \dot A - 2 x' C - x' B - x B' + 2 \dot x D = \alpha x C' - \alpha x \dot D, 
\end{equation}
where the terms depending on $w$ disappear. 
This has a simple solution for 
\begin{equation}
	\alpha = 2, \quad A = - 2 D, \quad B = -2C. 
	\label{solcon1} 
\end{equation}
We can use the same trick for the other constraint defining 
\begin{equation}
	\tilde y' +N z x' -N x z' = E, \quad y' = F, \quad \tilde z' +N x y' -N y x' = u' +N x G, \quad z' = \dot u+N x H, 
\end{equation}
so that (\ref{constz1}) becomes 
\begin{equation}
	\dot x G + x \dot G + 2 x' E - x' H - x H' - 2 \dot x F = - \alpha x E' + \alpha x \dot F. 
\end{equation}
The solution in this case is 
\begin{equation}
	\alpha = 2, \quad G = 2 F, \quad H = 2 E. 
	\label{solcon2} 
\end{equation}
Putting together the information coming from (\ref{solcon1}) and (\ref{solcon2}) we finally obtain the redefinitions of the various coordinates: 
\begin{equation}
	y' = \frac{\dot w - 2N x u'}{1+ 4N^2 x^2}, \qquad z' = \frac{\dot u + 2 N x w'}{1+ 4N^2 x^2}, 
\end{equation}
and 
\begin{equation}
	\tilde y' +N z x' -N x z' = \frac{w' - 2N x \dot u}{1+ 4N^2 x^2}, \qquad \tilde z' -N y x' +N x y' = \frac{u' + 2N x \dot w}{1+ 4N^2 x^2}, 
\end{equation}
or 
\begin{equation}
	\begin{array}{l}
		\dot w = (1+ 2N^2 x^2)y' + 2Nx \tilde z' - 2N^2 xyx', \\[2mm]
		\dot u = (1+ 2N^2 x^2)z' - 2Nx \tilde y' - 2N^2 xzx', \\[2mm]
		w' = (\tilde y +N x z)', \\[2mm]
		u' = (\tilde z -N x y)'. 
		\\[2mm]
	\end{array}
	\label{uwdef} 
\end{equation}
Note that $(\dot u)' = (u')\dot{}\ $ by using the constraint equation, modulo the $\tilde z$ equation of motion (and a similar argument holds for $w$). 
From this rewriting we explicitly see that the $w$ and $u$ coordinates are nothing but $\tilde y$ and $\tilde z$, shifted by a coordinate dependence on the dual ones. 
We can also use (\ref{uwdef}) to understand the behaviour of these coordinates under $x \to x + 1$ monodromies (they remain fixed under $z \to z+1$ and $y \to y+1$, while they obviously shift in the same way as $\tilde y$ and $\tilde z$). 
Whenever we shift $x$ by one period, we can explicitly see that the $u$ and $w$ coordinates have to obey non-local trasformations, namely: 
\begin{equation}
	\begin{array}{l}
		\displaystyle w' \to w' + 2 \frac{\dot u + 2N x w'}{1+ 4N^2 x^2}, \\[4mm]
		\displaystyle \dot w \to \dot w + 2 \frac{u' + 2N x \dot w}{1+ 4N^2 x^2}, \\[4mm]
		\displaystyle u' \to u' -2 \frac{\dot w - 2N x u'}{1+ 4N^2 x^2}, \\[4mm]
		\displaystyle \dot u \to \dot u -2 \frac{w' - 2N x \dot u}{1+ 4N^2 x^2}. 
	\end{array}
	\label{boundcon} 
\end{equation}
We will come back to these later on.

Using the (\ref{uwdef}) redefinitions we now get that the equations of motion for $\tilde y$ and $\tilde z$, or better, for $w$ and $u$ read 
\begin{equation}
	\left(\frac{w' - 2N x \dot u}{1+ 4N^2 x^2}\right)^\prime = \left(\frac{\dot w - 2N x u'}{1+ 4N^2 x^2}\right)^{{}^\bullet} 
\end{equation}
and 
\begin{equation}
	\left(\frac{u' + 2N x \dot w}{1+ 4N^2 x^2}\right)^\prime = \left(\frac{\dot u + 2N x w'}{1+ 4N^2 x^2}\right)^{{}^{\bullet}}, 
\end{equation}
which are the appropriate equations of motion for a $Q$-flux background as we will see in a moment. 
Using the same trick in the equation of motion for $x$ (in detail adding to the $x$ equation of motion $y$ times the $\tilde z$ equation of motion and subtracting $z$ times the $\tilde y$ equation of motion) one gets 
\begin{equation}
	0= \ddot x - x'' + \frac12 \,\frac{\partial\phantom{x}}{\partial x} \left[ \frac{\dot u^2+\dot w^2 - {u^\prime}^2-{w^\prime}^2 + 4 x N \left(\dot u w' - u' \dot w\right)}{1+ 4N^2 x^2}\right]. 
\end{equation}
Altogether these equations of motion are derived from a standard $\sigma$-model with a metric 
\begin{equation}
	ds^2 = dx^2 + \frac{1}{1+4N^2 x^2}\left(du^2 + dw^2\right), 
\end{equation}
and $B$-field 
\begin{equation}
	B = \frac{2Nx}{1+ 4N^2 x^2} du \wedge dw. 
\end{equation}

The problematic geometric interpretation of this background is not simply a consequence of the twisted boundary conditions but also of the field redefinition (\ref{uwdef}). 
One can see this by inspecting a simple string motion in the TDT and the resulting projection to this background. 
A very easy solution of the classical string equations is 
\begin{equation}
	x = \alpha \tau, \quad \tilde x = \alpha \sigma, \quad \tilde z = \beta \sigma, \quad y=z=\tilde y = 0. 
\end{equation}
This represents a string wrapped on the $\tilde x$ and $\tilde z$ coordinates and moving along $x$. 
All boundary conditions are trivially respected. 
After integrating out $\tilde x$, $y$ and $z$, the resulting configuration is given by a string wrapped around $u = \beta \sigma$, and not only moving in the $x$ direction, but also in $w$. 
This is forced by the field redefinition (\ref{uwdef}) and by the new twisted boundary conditions (\ref{boundcon}). 
These new conditions imply that when the string has moved by 1 in the $x$ direction, it must have moved also in the $w$ direction, since it is wrapped along $u$ and thus $u' \neq 0$. 
Actually the identification along $w$ for this motion is 
\begin{equation}
	\left\{ 
	\begin{array}{c}
		x\sim x+1 \\[2mm]
		w \sim w + 2 \beta/\alpha x 
	\end{array}
	\right. 
\end{equation}
and the motion in $w$ goes like $ w = \alpha \beta \tau^2$.

% subsection _q_flux (end)
\subsubsection{$R$-flux} % (fold)
\label{sub:_r_flux}

The last possible choice of coordinates leading to a non-geometric background is given by integrating out $x$, $y$ and $z$ and keeping only the coordinates dual to the original $H$-flux background, leading to the so-called $R$-flux configuration. 
Inspection of the doubled metric ${\cal H}$ reveals that one can indeed introduce a metric and a $B$-field for such a model, but that these fields depend explicitly on the dual coordinates \cite{Dall'Agata:2007sr}, which in this frame are $x,y$ and $z$.
The solution of the corresponding constraints will provide these coordinates as functions of the background coordinates $x = x(\tilde x, \tilde y, \tilde z)$, $y = y(\tilde x, \tilde y, \tilde z)$ and $z = z(\tilde x, \tilde y, \tilde z)$.
Unfortunately these constraints are not easy to solve for $x,y$ and $z$, since they are not first order constraints as the ones leading to the $h$ and $\tau$ flux.
In particular we expect that their solution leads to a non local expression for the $x,y$ and $z$ coordinates in terms of the dual ones, similarly to what happens for the flat group, when integrating out $x$.
This would explain the fact that this metric is not ``geometric'' even locally.

% subsection _r_flux (end)

% subsection the_h_tau_q_r_flux_chain (end)

\subsection{Chiral WZW models} % (fold)
\label{sub:chiral_wzw_models}

In this section we discuss the 6-dimensional TDT arising from the compact group SU(2) $\times$ SU(2).
There are two obvious embeddings in O(3,3), which, in the language of the previous section, can be described either by $\tau$ and $R$ fluxes, or by $Q$ and $H$. Note however, that these two choices are related by three T-dualities.
We choose to start from the $\tau$, $R$ algebra with structure constants
\begin{equation}
	\begin{array}{l}
	{\cal T}_{12}{}^3 = -1, \quad	{\cal T}_{32}{}^1 = -1, \quad 	{\cal T}_{23}{}^1 = -1,\\[2mm]
	{\cal T}^{12\,3} = -1, \quad	{\cal T}^{32\,1} = -1, \quad 	{\cal T}^{23\,1} = -1,\\[2mm]
	{\cal T}_1{}^2{}_3 = -1, \quad	{\cal T}_3{}^2{}_1 = -1, \quad 	{\cal T}_2{}^3{}_1 = -1.\\[2mm]
	\end{array}
\end{equation}
The TDT can be constructed in the usual way.
A convenient choice for the group element is the following:
\begin{equation}
	g = {\rm e}^{(Z_1+X^1)(x + \tilde x)}{\rm e}^{(Z_2+X^2)(y + \tilde y)}{\rm e}^{(Z_3+X^3)(z + \tilde z)}{\rm e}^{(Z_1-X^1)(x - \tilde x)}{\rm e}^{(Z_2-X^2)(y - \tilde y)}{\rm e}^{(Z_3-X^3)(z - \tilde z)}.
\end{equation}
This leads to the following background data: 
\begin{equation}
	{\cal H} = \left( 
	\begin{array}{cccccc}
		1 & &-\cos \tilde y \sin y  & & &-\cos y \sin \tilde y \\
		&1&&  & & \\
		-\cos \tilde y \sin y && 1& -\cos y \sin \tilde y&   & \\
		& & -\cos y \sin \tilde y& 1 & &-\cos \tilde y \sin y \\
		&&& & 1 & \\
		-\cos y \sin \tilde y & &&-\cos \tilde y \sin y & & 1 \\
	\end{array}
	\right)
\end{equation}
and 
\begin{equation}
	\eta = \left( 
	\begin{array}{cccccc}
		  & &-\cos y \sin \tilde y  & 1& & -\cos \tilde y \sin y\\
		& &&  & 1& \\
		-\cos y \sin \tilde y &&  & -\cos \tilde y \sin y&   &1 \\
		1& & -\cos \tilde y \sin y&   & &-\cos y \sin \tilde y \\
		&1&& &   & \\
		 -\cos \tilde y \sin y& &1&-\cos y \sin \tilde y & &   \\
	\end{array}
	\right).
\end{equation}
Note that $\eta$ cannot be put in a constant form because the TDT is the $S^3 \times S^3$ group manifold.
It should be noted that the structure of the matrices has a common pattern
\begin{equation}
	{\cal H} = \left(\begin{array}{cc}
	 {\cal A}&{\cal B}\\ {\cal B}& {\cal A} 
	\end{array}\right), \quad \eta = \left(\begin{array}{cc}
{\cal B}& {\cal A} \\ {\cal A} & {\cal B}
	\end{array}\right).
\end{equation}
We will come back on the explanation of this form later on.

The standard TDT construction defines also the 2-form $C$ in a specific way from (\ref{dC}), but we have seen that for a compact group the opposite sign choice leads to first order equations.
Following this route, we get that
\begin{equation}
	C = - \sin \tilde y \cos y (dx \wedge dz + d\tilde x \wedge d\tilde z) - \sin y \cos \tilde y (dx \wedge d\tilde z + d\tilde x \wedge dz).
\end{equation}
Such a choice reduces the Lagrangian to a very simple form.
Using the chiral basis $y^i = \frac12(y^i_L + y^i_R)$, $\tilde y_i = \frac12(y^i_L - y^i_R)$ this reads (here $\partial_\pm = \partial_0 \pm \partial_1$)
\begin{equation}
	{\cal L} = \frac12\, \partial_+ y^i_R\, \partial_1 y^j_R\, {\mathbb C}^R_{ij} - \frac12\, \partial_- y^i_L\, \partial_1 y^j_R\, {\mathbb C}_{ij}^L,
\end{equation}
where
\begin{equation}
	{\mathbb C}^L = \left(\begin{array}{ccc}
	 1 &  & -2 \sin y_L \\ & 1& \\ && 1 
	\end{array}\right)
\end{equation}
and
\begin{equation}
	{\mathbb C}^R = \left(\begin{array}{ccc}
	 1 &  & -2 \sin y_R \\ & 1& \\ && 1 
	\end{array}\right).
\end{equation}
This Lagrangian has the interpretation of the sum a chiral and an antichiral WZW model on the SU(2) group manifold \cite{Tseytlin:1990va} with ${\mathbb C}^{L,R} = g^{L,R} + B^{L,R}$, for the SU(2) metric
\begin{equation}
	ds^2 = (dy_1 - \sin y_2 d y_3)^2 + dy_2^2 + \cos^2 y_2 dy_3^2
\end{equation}
and the corresponding B-field
\begin{equation}
	B = -\sin y_2 dy_1 \wedge dy_3,
\end{equation}
chosen so that $dB$ is the volume form. 
It is therefore clear that the first order equations of motion for this model, $V_I =0$, are equivalent to the equations of motion of the (anti) chiral fields.

From this analysis it is also now visible that the general structure of the ${\cal H}$ and $\eta$ matrices for a TDT that is the product of two compact gauge groups ${\cal G}_1 \times {\cal G}_2 \subset {\rm O}(d) \times {\rm O}(d) \subset {\rm O}(d,d)$ has to follow the pattern outlined above, because  
\begin{equation}
	{\cal H} = \left(\begin{array}{cc}
	 g_R+g_L&g_R-g_L\\ g_R-g_L& g_R+g_L 
	\end{array}\right), \quad \eta = \left(\begin{array}{cc}
g_R-g_L& g_R+g_L \\ g_R+g_L & g_R-g_L
	\end{array}\right).
	\label{HgLgR}
\end{equation}
Also the structure of the antisymmetric form $C$ is related to $B_L$ and $B_R$ following the same pattern.
In general if ${\cal G}_1 \neq {\cal G}_2$ this model is not equivalent to a standard WZW theory and it is tempting to think of it as a non-geometric generalization of the WZW model.

From this example we can also learn that there is no clear universal recipe to extract from the TDT data the information regarding the metric and $B$ field of the string $\sigma$-model obtained by integrating out half of the coordinates.
For instance, these data can be completely contained in ${\cal H}$ in a non-linear way as in (\ref{HgB}), but the same matrix may contain only the information on the metric, as in the last example (\ref{HgLgR}).

% subsection chiral_wzw_models (end)

% section examples (end)

\bigskip

\bigskip

\section*{Acknowledgments.}

\noindent We are glad to thank K.~Lechner, K.~Sfetsos and D.~Sorokin for discussions. 
The research is supported by the European Union under the contract MRTN-CT-2004-005104, ``Constituents, Fundamental Forces and Symmetries of the Universe'' in which G.~D.~is associated to Padova University.

\bibliographystyle{plain}

\end{document}